**Title page**

**Classification:** PHYSICAL SCIENCE: Physics

**Title:** Superconductive "sodalite"-like clathrate calcium hydride at high pressures

**Author affiliation:** Hui Wang[1], John S. Tse[2], Kaori Tanaka[2], Toshiaki Iitaka[3], and Yanming Ma[1*]

[1]State Key Lab of Superhard Materials, Jilin University, Changchun 130012, P. R. China,

[2]Department of Physics and Engineering Physics, University of Saskatchewan, Saskatoon, Saskatchewan, S7N 5E2, Canada,

[3]Computational Astrophysics Laboratory, RIKEN, 2-1 Hirosawa, Wako, Saitama 351-0198, Japan.

**\*Corresponding author:**

Yanming Ma

*State Key Lab of Superhard Materials, Jilin University, Changchun 130012, China*

TEL: +86-431-85168276
FAX: +86-431-85168276
EMAIL: mym@jlu.edu.cn

**Manuscript information:** 9 text pages and 4 figures

**Abbreviations:** Calcium hydride | High pressure | sodalite structure

# Abstract

**Hydrogen-rich compounds hold promise as high-temperature superconductors under high pressures. Recent theoretical hydride structures on achieving high-pressure superconductivity are composed mainly of $H_2$ fragments. Through a systematic investigation of Ca hydrides with different hydrogen contents using particle-swam optimization structural search, we show that in the stoichiometry $CaH_6$ a body-centred cubic structure with hydrogen that forms unusual "sodalite" cages containing enclathrated Ca stabilizes above pressure 150 GPa. The stability of this structure is derived from the acceptance by two $H_2$ of electrons donated by Ca forming a "$H_4$" unit as the building block in the construction of the 3-dimensional sodalite cage. This unique structure has a partial occupation of the degenerated orbitals at the zone centre. The resultant dynamic Jahn–Teller effect helps to enhance electron–phonon coupling and leads to superconductivity of $CaH_6$. A superconducting critical temperature ($T_c$) of 220–235 K at 150 GPa obtained from the solution of the Eliashberg equations is the highest among all hydrides studied thus far.**

\body

**Introduction**

Current studies of the hydrides of simple elements at high pressures are motivated by the proposition that a high $T_c$ may be achieved from "pre-compressed" $H_2$ in dense hydrogen-dominant hydrides[1]. Superconductivity has been reported in $SiH_4$[2], although controversy surrounds models of the origin of superconductivity and the structure of the superconducting state[3-5]. Many theoretical structures have the predicted $T_c$ values ranging from 10 to 139 K[6-12]. Except for $AlH_3$[13], in which superconductivity was predicted but not observed, a common feature of these structures is the ubiquitous presence of molecular hydrogen fragments. At high pressures, unfavourable electron–electron repulsion in the atomic valence shells can be reduced by migration and localization in the empty interstitial regions[14,15]. Introduction of an electronegative element, as was demonstrated in the case of K-Ag[16], increases the ability of the electronegative Ag atoms to accommodate electrons from K into the outermost $5s$ and $5p$ valence shells, thereby forming a bonding network and stabilizing the new alloy structures that otherwise would not be observed under ambient pressures. Recently, Zurek *et al.*[17,18] demonstrated that charge transfer from lithium or sodium to hydrogen molecules under high pressures could lead to the formation of new metallic lithium- or sodium-hydride alloys. Depending on the stoichiometry, the structures consisting of "pre-dissociated" molecular $H_2$ and/or monoatomic hydrogen.

**Results and Discussion**

The known calcium hydrides have a stoichiometry of $CaH_2$ at ambient pressures. Here, we explored other calcium hydrides with larger hydrogen contents by compressing a mixture containing Ca + hydrogen or $CaH_2$ + hydrogen. The search for stable $CaH_{2n}$ (n = 2 – 6) structures at high pressures was performed using the particle-swam optimization structure prediction algorithm[19] in combination with *ab initio* calculations. For each stoichiometry, calculations were performed at pressures 50-200 GPa with up to four formula units in the model. The enthalpies of candidate structures relative to the products of dissociation into $CaH_2$ + solid $H_2$ at the appropriate pressures are summarized in Fig. 1(a). The essential information can be summarized as follows: (i) except for $CaH_{10}$, stable structures began to emerge at pressures < 50 GPa; (ii) $CaH_4$ was the most stable phase at pressures between 50 and 150 GPa, while at 200 GPa $CaH_6$ had the lowest enthalpy of formation; (iii) the breakup of hydrogen molecules depended on the Ca/H ratio, and a higher susceptibility to dissociation was observed at higher ratios. Notably, calcium hydrides with stoichiometry having odd number of hydrogen (e.g. $CaH_3$, $CaH_5$, and $CaH_7$, etc) were found to be energetically very unfavourable and were excluded in the discussions (see Supplementary Information).

The structures of the stable phases for each stoichiometry at 150 GPa are shown in Fig. 1(b)-(d). Three types of hydrogen species, "$H_4$" units, monatomic H + $H_2$, and molecular $H_2$, were observed. $CaH_4$ had a tetragonal ($I4/mmm$, Pearson symbol tI10) structure and included a body-centred arrangement of Ca and two molecular and four monatomic hydrogen units ($Ca_2(H_2)_2(H)_4$). The structure of $CaH_6$ adopted a remarkable cubic form ($Im\bar{3}m$, Pearson symbol cI14), with body-centred Ca atoms and, on each face, squared "$H_4$" units tilted 45° with respect to the plane of the Ca atoms. These "$H_4$" units were interlinked to form a sodalite framework with a Ca atom enclathrated at the centre of each cage. The next stable polymorph, $CaH_{12}$, had a rhombohedral ($R\bar{3}$, Pearson symbol hR13) structure consisting entirely of molecular $H_2$.

The presence of different types of hydrogen can be rationalized based on the effective number of electrons contributed by the Ca atom and accepted by each $H_2$ molecule. Assuming that the two valence electrons of each Ca atom were completely "ionized" and accepted by $H_2$ molecules, the "formal" effectively added electron (EAE) per $H_2$ for $CaH_4$ was $1e/H_2$, for $CaH_6$ was $(2/3)e/H_2$, and was $(1/3)e/H_2$ for $CaH_{12}$. Because $H_2$ already had a filled σ bond, the added electrons resided in the antibonding σ* orbitals, which weakened the H-H bond (i.e., lengthened the H-H bond length) and eventually resulted in complete dissociation. The presence of H and $H_2$ units in the $CaH_{2n}$ structures depended on the number of EAEs. Two formulae were present in each unit cell of $CaH_4$. Because half (two) of each $H_2$ molecule was retained, the remaining two $H_2$ molecules had to accommodate four "excess" electrons into their $\sigma^*$ orbital, which broke up the molecules into monatomic hydrides ($H^-$). The formation of "$H_4$" units in $CaH_6$ was not accidental. If molecular hydrogen atoms were present, each $\sigma^*$ orbital of $H_2$ had to accommodate $(2/3)e$. This led to a physically unfavourable structure with significant weakening of the intramolecular H-H bonds. There is, however, an alternative lower energy structure, the one predicted here. From molecular orbital theory, a square "$H_4$" unit should possess a half-filled degenerated "non-bonding" orbital (Fig. 2a). This orbital, in principle, can accommodate up to two electrons without detrimentally weakening the H-H bond (Fig. 2b). It should be noted that at 150 GPa, the H…H distance of "$H_4$" in the cI14 structure of $CaH_6$ was 1.24 Å, which was substantially shorter than both the monatomic H…H distance of 1.95 Å and the H…$H_2$ distance of 1.61 Å in $CaH_4$, but in good agreement with those of 1.27 and 1.17 Å in isolated $H_4^{2-}$ and "$H_4$" squares, respectively. This suggested the presence of a weak covalent H…H interaction in $CaH_6$. As will be described below, the "$H_4$" unit is the fundamental building block of the sodalite cage.

In $CaH_{12}$, the EAE was $(1/3)e/H_2$, which could be taken up by $H_2$ without severing the bond. Extending this concept further then predicted that $CaH_{12}$ and hydride-alloys with a high H content (smaller EAE) would be composed predominantly of molecular $H_2$. An important observation in support of the EAE description given above is that the H-H bond in $CaH_4$ lengthened from 0.81 Å at 100 GPa to 0.82 Å at 150 GPa, while the H-H bond in $CaH_{12}$ shortened from 0.81 Å to 0.80 Å. Here, more Ca valence electrons in $CaH_4$ were available for transfer to the $H_2$ $\sigma^*$ orbitals than were available in $CaH_{12}$, thereby more severely weakening the bond in $CaH_4$. This effect was relatively small for $CaH_{12}$, in which the H-H bond was shortened due to compression.

The zero-point motion was not included in the calculation of the formation enthalpy of the various hydrides (Fig. 1a), although it is expected to be very influential due to the presence of large amounts of hydrogen. We estimated the zero-point energies of $CaH_6$ and $CaH_4$ using the quasiharmonic model[20] at 150 GPa. It was found that the inclusion of zero-point motion significantly lowered the formation enthalpy of $CaH_6$ with respect to $CaH_4$ (Supplementary Figure S11). As a consequence, $CaH_6$ became more stable at and above 150 GPa. The physical mechanism underlying this effect stemmed from the "$H_4$" moieties, which included much longer H-H distances and led to significantly softened phonons. This contrasted with other Ca hydrides studied (e.g., $CaH_4$ and $CaH_{12}$), in which the presence of "$H_2$" molecular units gave rise to higher frequency phonons and, thus, a larger zero-point energy.

The three-dimentional sodalite cage in $CaH_6$ is the result of interlink of other "$H_4$" units via each H atom at the corner of one "$H_4$" unit. So, what is the electronic factor promoting the formation of these "$H_4$" units? To answer this question, the electron localization functions (ELF) of a hypothetical bare bcc Ca lattice with the H atoms removed and $CaH_6$ hydride (Fig. 3a and 3b) were examined. In bare bcc Ca, regions with ELF values of 0.58 were found to localize at the H atom sites in the "$H_4$" units on the faces of the cube. The ELF of $CaH_6$ hydride suggested that no bonds were present between the Ca and H. A weak "pairing" covalent interaction with an ELF of 0.61, however, was found between the H atoms that formed a square "$H_4$" lattice. Their formation resulted from the accommodation by $H_2$ of "excess" electrons from the Ca. An electron topological analysis also showed the presence of a bond-critical point[21] along the path connecting neighbouring H atoms. The integrated charge within the H atomic basin was 1.17$e$, which corresponded to a charge transfer of 1.02$e$ from each Ca. A partially "ionized" Ca was also clearly supported by the band structure and the density of states as reported in Fig. 3c and Supplementary Fig. S15. At 150 GPa, Ca underwent an *s-d* hybridization with an electron transferred from the 4*s* to the 3*d* orbital. In $CaH_6$, the Ca site symmetry was $m\bar{3}m$ ($O_h$) and the Ca 3$d$ manifold was clearly split into the $e_g$ and $t_{2g}$ bands, with the lower energy $e_g$ band partially occupied.

A comparison of the band structures of $CaH_6$, "$H_6$" ($Ca_0H_6$), and bare Ca ($CaH_0$) provided additional supporting evidence. Even without the presence of Ca, the valence band width of the hypothetical "$H_6$" (Fig. 3d) was 15.2 eV, comparable to 16.4 eV for $CaH_6$ (Fig. 3c). In comparison, the valence band width of the "bare" bcc Ca was only 4.3 eV (Fig. 3e). The band structure of $CaH_6$ near the Fermi level was modified from "$H_6$" due to the the hybridization between Ca 3$d$ and H 1$s$ orbital; however, the trend in the electronic band dispersions from –2 to –16.4 eV was remarkably similar to that of "$H_6$" from 3 to –15.2 eV.

Sodalite cage was constructed from linking of "$H_4$" units and this topologically resulted in the formation of "$H_6$" faces. In fact, a primitive cell of $CaH_6$ can be seen as composed of a "$H_6$" hexagon and a Ca. Band structure analysis suggested that four conduction bands of "$Ca_0H_6$" sodalite cage (Fig. 3d) were half-filled and the addition of maximal two electrons to "$H_6$" gave rise to partial occupancy of the degenerate bands at Γ as indicated in $CaH_6$ (Fig. 3c). Partial occupancy of a degenerate orbital results in an orbitally degenerate state and is subject to Jahn–Teller (JT) distortion (Fig. 2)[22]. The JT effect involves coupling between the electron and nuclear degrees of freedom, leading to distortions in the structure, and it lifts the orbital degeneracy. If the distortion is dynamic, JT vibrations can contribute to superconductivity. The same mechanism has been invoked to explain the superconductivity in B-doped diamond[23]. To investigate this possibility, electron–phonon coupling (EPC) calculations on the sodalite structure of $CaH_6$ at 150 GPa were performed. The phonon dispersion curves, phonon linewidth $\gamma(\omega)$, EPC parameter $\lambda$, and Eliashberg spectral function $\alpha^2F(\omega)$ were calculated. A gap at 430 cm$^{-1}$ (Fig. 4) separated the phonon spectrum into two regions: the lower frequency branches were associated with the motions of both Ca and H, whereas the higher frequency branches were mainly associated with H atoms. The combined contribution (19% and 81%, respectively) gave an EPC parameter λ of 2.69. The calculated phonon linewidths (Fig. 4) showed that the EPC was derived primarily from the $T_{2g}$ and $E_g$ modes at the zone centre Γ. Incidentally, these two bands, respectively, were the in-plane breathing and rocking vibrations of the H atoms belonging to the "$H_4$" unit. The corresponding atomic vibrations led to distortions in the square planar structure. Moreover, both phonon branches showed significant phonon softening along all symmetric directions. A large EPC also benefited from the high density of states at the Fermi level caused by a Van Hove singularity at Γ (Fig. 3c). The very large EPC was unprecedented for the main group hydrides. Previous calculations on a variety of systems predicted an EPC in the range of 0.5–1.6[11]. The mechanism suggested here is not inconsistent with the mechanisms found in JT-induced superconductivity in alkali intercalated $C_{60}$, in which the intramolecular vibrations are responsible for distortions that lower the symmetry of the molecules, which is favourable for electron–phonon processes[24,25].

$T_c$ was calculated based on the spectral function $α^2F(ω)$ by numerically solving the Eliashberg equations[26], which consist of coupled non-linear equations describing the frequency-dependent order parameter and renormalisation factor. The Coulomb repulsion is taken into account in terms of the Coulomb pseudopotential, μ*, scaled to a cutoff frequency (typically six times the maximum phonon frequency)[27]. At 150 GPa, the predicted $T_c$ values were 235 K and 220 K using typical values for μ* of 0.1 and 0.13, respectively. EPC calculations were also performed for 200 GPa and 250 GPa, in which the calculated $T_c$ was found to decrease with pressure (201 K at 200 GPa and 187 K at 250 GPa for μ*= 0.13), with a pressure coefficient ($dT_c/dP$) of –0.33 K/GPa. $T_c$ of the order of 200 K is among the highest for all reported hydrides.

The predicted high $T_c$ for $CaH_6$ is very encouraging, but it must be viewed with caution. Formally, there is no upper limit to the value of $T_c$ within the Midgal–Eliashberg theory of superconductivity. Two practical factors must be considered. The calculation of the EPC is based on the harmonic approximation and without consideration of electron correlation effects. Because strong electron–phonon coupling in $CaH_6$ arises from the proximity of the electronic and structural instabilities, anharmonicity of the atomic motions can lead to renormalization of the vibrational modes, as demonstrated in a study of $AlH_3$. In that study, lower renormalized frequencies were found to reduce the EPC and suppress superconductivity[28]. On the other hand, the electron–phonon matrix elements may be enhanced by anharmonic vibrations, as in the case of disordered materials[29]. In a recent hybrid functional study of $C_{60}$ anions, the inclusion of Hartree–Fock exchange contributions was shown to have little effect on the structural properties and phonon frequencies, but resulted in a strong increase in the electron–phonon coupling[24].

The formation of a hydrogen sodalite cage with enclathrated calcium in $CaH_6$, reported here for hydrogen-rich compounds, provides an unexpected example of a good superconductor created by the compression of a mixture of elemental calcium + hydrogen or $CaH_2$ + hydrogen. This novel superconductor can also be viewed as consisting of unique square "$H_4$" units and electron-donating calcium atoms subject to JT effects. Dense superconductive states, such as those reported here, may be favoured in other mixtures of elemental metals + hydrogen or any hydride + hydrogen upon compression. This work highlights the major role played by pressure in effectively overcoming the kinetic barrier to formation in the synthesis of novel hydrides.

**Methods**

Our structure prediction approach is based on a global minimization of free energy surfaces merging *ab* initio total-energy calculations via particle swarm optimization technique as implemented in CALYPSO (Crystal structure AnaLYsis by Particle Swarm Optimization) code[19]. Our CALYPSO method unbiased by any known structural information has been benchmarked on various known systems[19] with various chemical bondings and had several successful prediction of high pressure structures of Li, Mg, and $Bi_2Te_3$[30-32], among which the insulating orthorhombic (*Aba*2, Pearson symbol oC40) structure of Li and the two low-pressure monoclinic structures of $Bi_2Te_3$ have been confirmed by independent experiments[32,33]. The underlying *ab initio* structural relaxations were carried out using density functional theory within the Perdew-Burke-Ernzerhof exchange-correlation[34] as implemented in the VASP code[35]. The all-electron projector-augmented wave method[36] was adopted with 1$s$ and 3$p^6$4$s^2$ treated as valence electrons for H and Ca, respectively. Electronic properties, lattice dynamics and electron-phonon coupling were studied by density functional (linear-response) theory as implemented in the QUANTUM ESPRESSO package[37]. More computational details can be found in the supplementary information.

**Acknowledgements**

H. W. and Y. M. acknowledge Prof. Aitor Bergara for the valuable discussions and are thankful to the financial supports by Natural Science Foundation of China (NSFC) under No. 11104104, the China 973 Program (No. 2011CB808200), NSFC under Nos. 11025418 and 91022029, and the research fund of Key Laboratory of Surface Physics and Chemistry (No. SPC201103) T. I. was supported by MEXT of Japan (No. 20103001-20103005). The calculations were performed in the computing facilities at RICC system in RIKEN (Japan) and the High Performance Computing Center of Jilin University.

**Author contributions**

Y.M. proposed the concept. Y. M. and H. W. designed the simulations. H. W., J. T., K. T., Y.I., and Y. M. carried out the simulations and conducted the data analysis. H. W., J. T., and Y. M. wrote the manuscript.

*Correspondence and requests for materials should be addressed to Y.M. (mym@jlu.edu.cn).

**Figure captions**

**Figure 1 | Enthalpies of formation (ΔH, with respect to CaH$_2$ and H$_2$) of CaH$_{2n}$ (n=2–6) and crystal structures. a.** The abscissa x is the fraction of H$_2$ in the structures. The open, solid, and half-filled symbols indicate that the structures are composed of H$_4$ units, molecular H$_2$, and the coexistence of H$_2$ and H, respectively. The metastable structures are indicated by circles. The stable pressure ranges for CaH$_4$, CaH$_6$, and CaH$_{12}$ are 50–200 GPa, 150–200 GPa, and 100–200 GPa, respectively. The EAE (per H$_2$) is shown in brackets. The estimated stability fields were determined according to the static enthalpies and may shift upon inclusion of dynamic effects (the zero-point motion of the nuclei). **b.** Structure of tI10-CaH$_4$. **c.** Structure of cI14-CaH$_6$. **d.** Structure of hR13-CaH$_{12}$. Monatomic H, molecular H$_2$, and Ca atoms are shown as cyan, green, and royal blue, respectively. The green cylinders and grey dashed lines are drawn to represent molecular H$_2$ and the sodalite cage, respectively.

**Figure 2 | Hückel energy-level diagrams of H$_4$ and H$_4^{2-}$ units. a.** Hückel energy-level diagram of H$_4$. **b.** Hückel energy-level diagram of H$_4^{2-}$. The partial occupation of electrons on the degenerate orbitals of H$_4$ units can lead to a Jahn–Teller distortion, but the H$_4^{2-}$ possesses a closed shell electronic structure and therefore no Jahn–Teller distortion is expected.

**Figure 3 | Electron localization function (ELF) and band structure. a.** ELF of CaH$_0$ (cI14 structure with H removed). **b.** ELF of CaH$_6$. **c.** Band structure of CaH$_6$. **d.** Band structure of Ca$_0$H$_6$ (cI14 structure with Ca removed). **c.** Band structure of CaH$_0$. The horizontal dotted lines indicate the Fermi level.

**Figure 4 | Phonon band structure and Eliashberg spectral function.** Phonon dispersion curves of cI14 at 150 GPa (left panel). Olive circles indicate the phonon

linewidth with a radius proportional to the strength. Phonons with a larger linewidth at Γ belong to the $t_{2g}$ and $e_g$ modes, as indicated by the circles at 960 cm$^{-1}$ and 1960 cm$^{-1}$, respectively. Eliashberg electron-phonon coupling spectral function $\alpha^2 F(\omega)$ at 150 GPa (right panel). Dashed line is the integration of the electron-phonon coupling strength as a function of phonon frequency. The horizon lines are drawn as a guide.

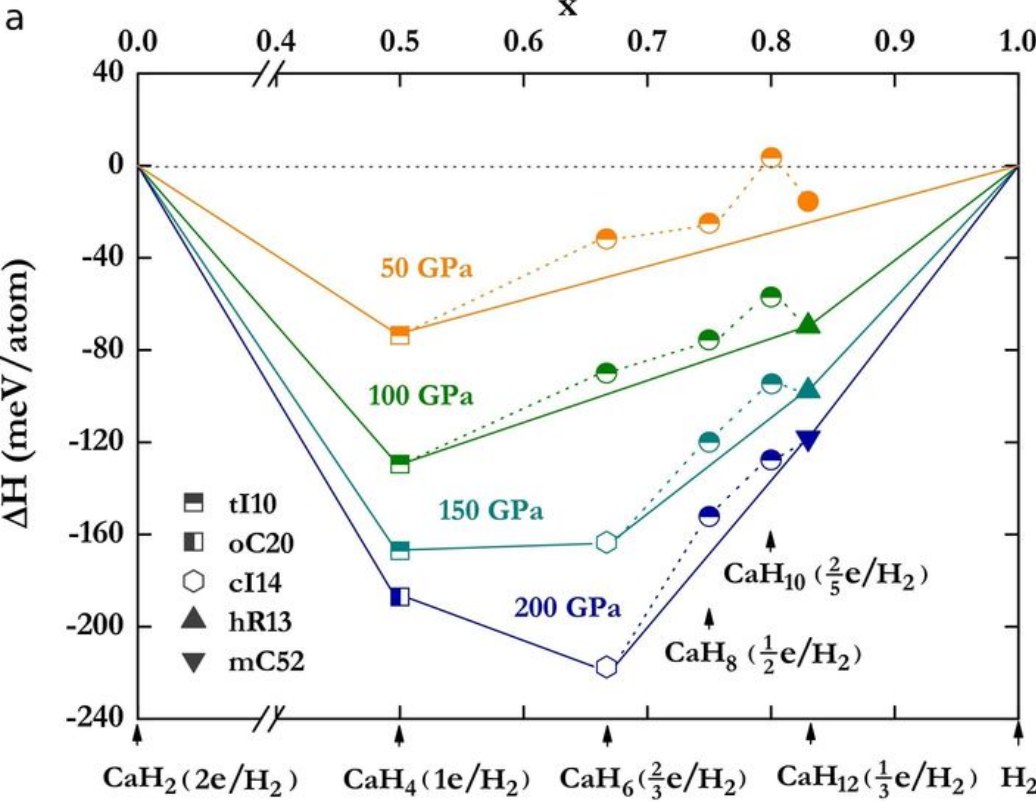

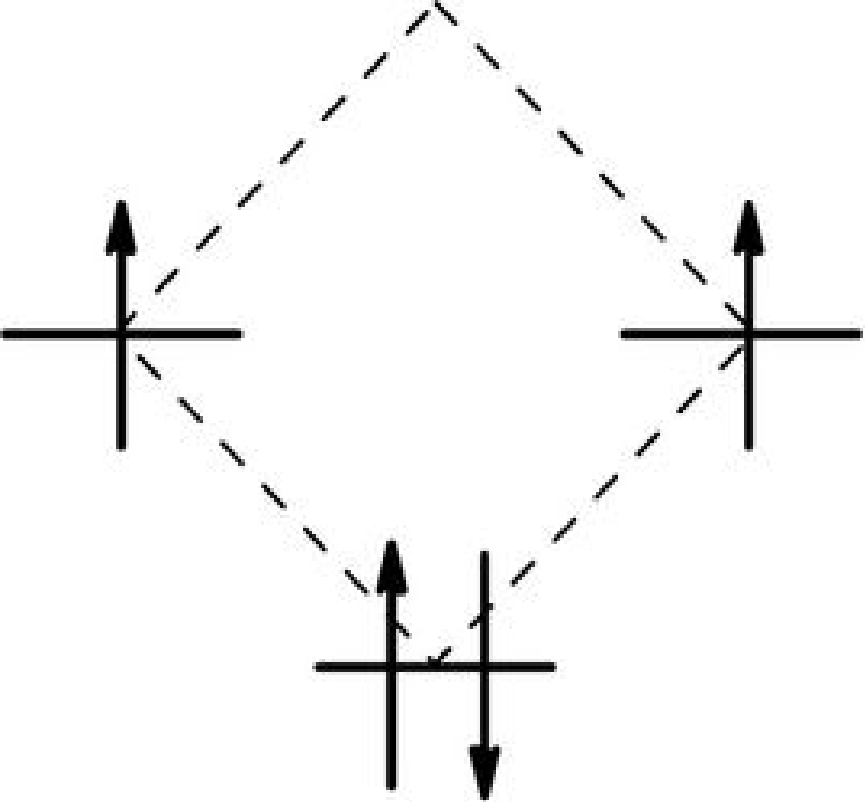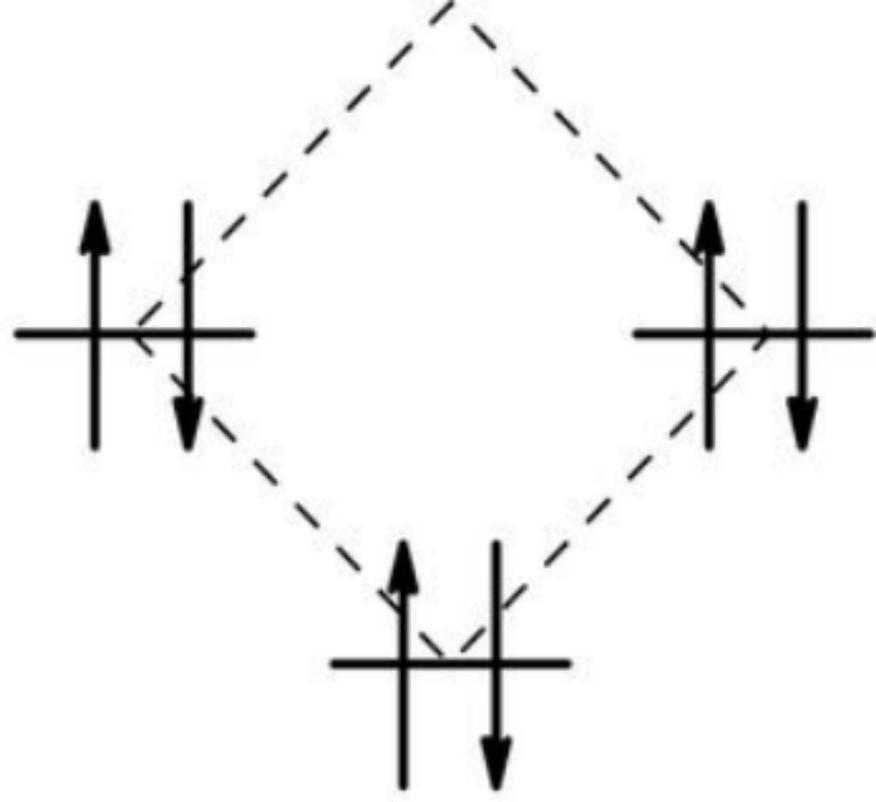

a    b

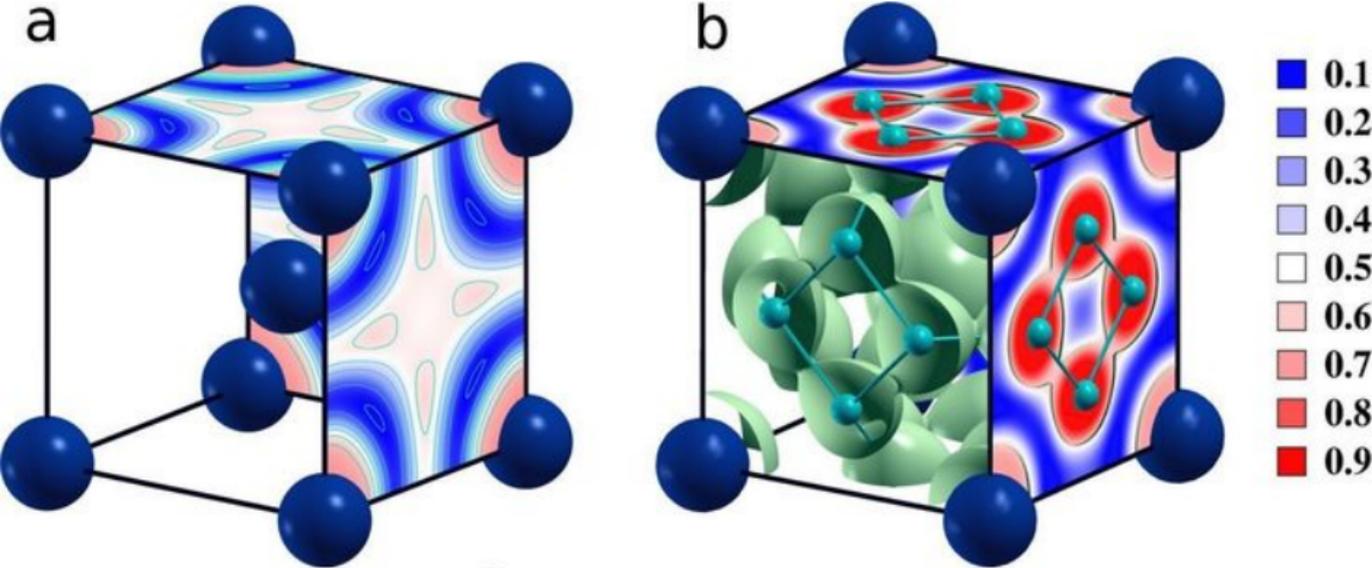

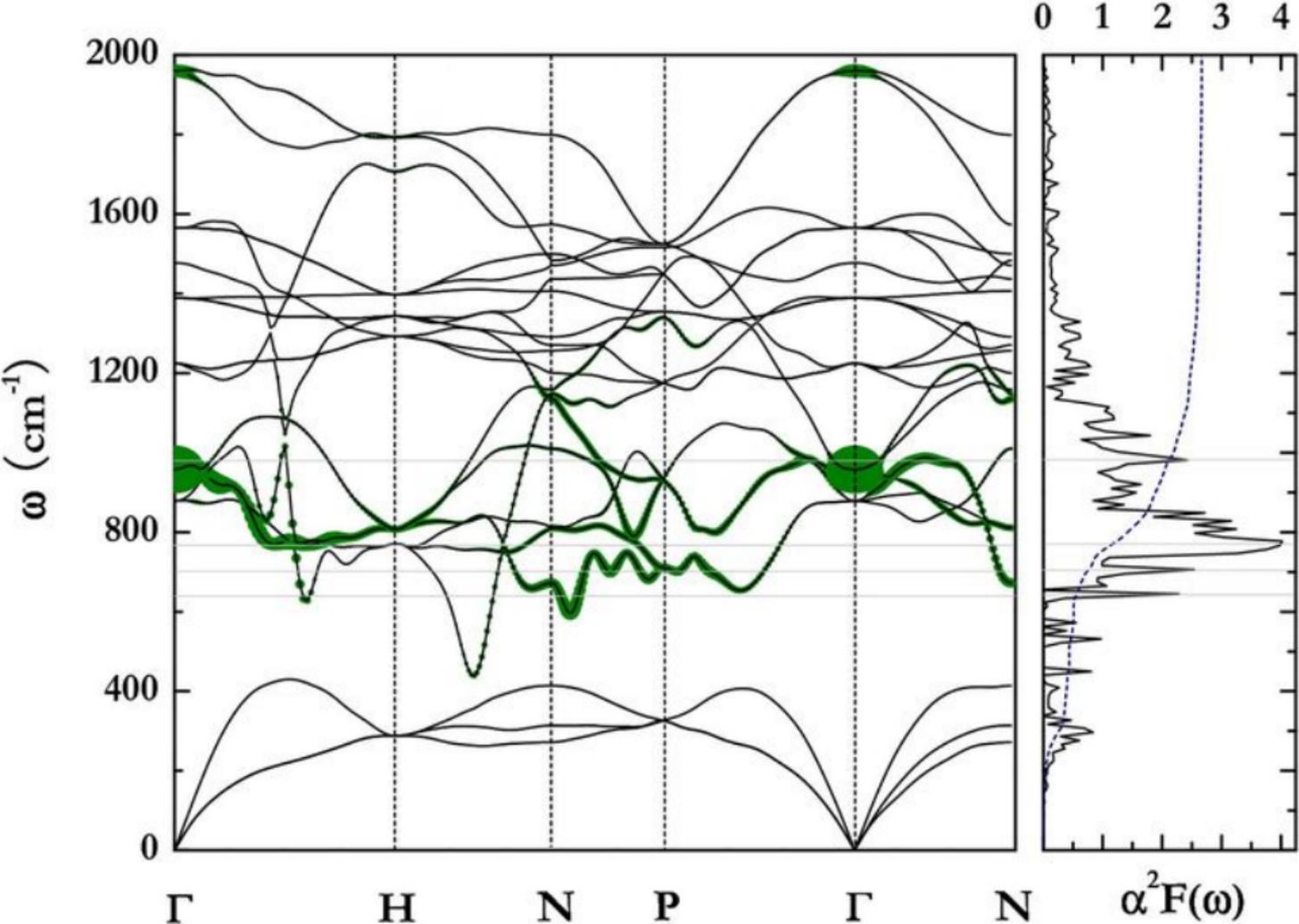